\documentclass[aps,pra,twocolumn,showpacs,groupedaddress]{revtex4}
\usepackage{graphicx}
\usepackage{amsmath}
\usepackage{longtable}

\begin{document}

\title{The higher order $C_{n}$ dispersion coefficients for \\
the alkali atoms.}
\author{J.Mitroy}  
\email{jxm107@rsphysse.anu.edu.au}
\author{M.W.J.Bromley}  
\email{mbromley@cdu.edu.au}
\affiliation{Faculty of Technology, Charles Darwin University, 
Darwin NT 0909, Australia}

\date{\today}

\begin{abstract}

The van der Waals coefficients, from $C_{11}$ through to 
$C_{16}$ resulting from 2nd, 3rd and 4th order perturbation 
theory are estimated for the alkali (Li, Na, K and Rb) atoms.  
The dispersion coefficients are also computed for all possible 
combinations of the alkali atoms and hydrogen.  The parameters 
are determined from sum-rules after diagonalizing the fixed 
core Hamiltonian in a large basis.  Comparisons of the radial 
dependence of the $C_n/r^n$ potentials give guidance as to the 
radial regions in which the various higher-order terms can be 
neglected.   It is seen that including terms up to 
$C_{10}/r^{10}$ results in a dispersion interaction that is 
accurate to better than $1\%$ whenever the inter-nuclear 
spacing is larger than 20 $a_0$.  This level of accuracy
is mainly achieved due to the fortuitous cancellation between 
the repulsive $(C_{11}, C_{13}, C_{15})$ and attractive 
$(C_{12}, C_{14}, C_{16})$ dispersion forces.    

\end{abstract}


\maketitle

\section{Introduction} 

The experimental realization of Bose-Einstein condensation 
(BEC) for the alkali atoms, Li, Na, Rb and atomic hydrogen 
\cite{ignuscio99a,leggett01a} has resulted in an upsurge of 
interest in the area of cold atom physics.  One consequence
of this is the increased importance in determining the
interaction potentials between alkali and alkaline-earth atoms.
For example, the stability and structure of BECs depends 
on the sign (and magnitude) of the scattering length, and the 
scattering length sensitively on the fine details of the 
interaction potential \cite{weiner99a,tiesinga02a}. 

One can try and determine the interaction potential by either 
explicit calculation \cite{spelsberg93a,geum01a,hinde03a}, or 
by analysis of high precision spectroscopic experiments of 
the dimer \cite{wang00a,vankempen02a,allard03a}.  These 
approaches are most suited to determining the 
potential at small to intermediate values of internuclear 
distance (e.g.  $r < 25$ $a_0$). At large distances, one 
relies on the approach pioneered by Dalgarno and collaborators 
\cite{dalgarno61a,dalgarno63a,dalgarno67a}, namely using 
oscillator strength sum rules to determine the so-called 
dispersion coefficients.               

The long-range interaction between two spherically 
symmetric atoms can be written in the general form 
\begin{equation}
V(r\to\infty) =  -V_6(r) - V_8(r) - V_{10}(r)    
 - V_{11}(r)  - V_{12}(r)  - \ldots ,  
\label{vdw1}
\end{equation}
where the dispersion potential, $V_n(r)$ of order $n$ is
written  
\begin{equation}
V_n(r) = \frac{C_n}{r^n}   \ . 
\label{vdw2}
\end{equation}
The $C_n$ parameters are the van der Waals dispersion 
coefficients.  The even ($n=6,8,\ldots$) dispersion coefficients are
calculated using oscillator-strength sum rules derived from
2nd-order perturbation theory and provide an attractive
interaction.  The odd ($n=11,13,\ldots$) terms come from
3rd-order perturbation theory, and are repulsive 
\cite{ovsiannikov88a}.  Contributions from 4th-order 
perturbation theory start at $n = 12$ \cite{bukta74a,ovsiannikov88a}.  
The calculations were stopped at $n = 16$ since the contributions 
from 5th and 6th-order perturbation theory start at $n = 17$ and 
$n = 18$ respectively.  

It has been customary in previous 
systematic studies of dispersion coefficients to restrict the 
calculations to just the $C_6$, $C_8$ and $C_{10}$ terms 
\cite{standard85a}.  
With the recent interest in determining the 
scattering lengths of A-A systems to high precision 
\cite{tiesinga02a,crubellier99a,venturi01a,vankempen02a,weiss04a,weiss04b}, 
it is now worthwhile to evaluate the 
higher $n$ dispersion coefficients with  a view to determining 
whether these make a significant contribution to the inter-atomic 
potential for those radial separations at which the long-range 
form of the potential is first applied. 

The recent spectrum analyses have typically an inter-nuclear
distance of about 20 $a_0$ as the boundary point used to join
the explicitly determined potential with the asymptotic form
given in terms of the dispersion forces 
\cite{wang00a,vankempen02a,allard03a}.   Taking the Rb-Rb 
potential as a specific example, explicit calculations have 
shown that $C_6 = 4.635 \times 10^3 $ au, 
$C_8 = 5.701 \times 10^5 $ au, $C_{10} = 7.916 \times 10^7 $ au,
and $C_{12} = 1.427 \times 10^{10} $ au \cite{mitroy03f}.  At
a radius of 20 $a_0$ the ratios of the contributions from the
various terms are $1:0.308:0.1068:0.0481$.  The contribution 
of the highest multipole potential is about $5\%$ of the 
dipole-dipole $C_6$ term.  
  
In this article, the $C_{11}$ to $C_{16}$ dispersion coefficients 
and related parameters are computed using a semi-empirical approach 
for the alkali atoms.  The method adopted utilizes oscillator strengths 
calculated within a semi-empirical framework \cite{mitroy03f}.  
Comparisons with previous high accuracy ab-initio calculations 
\cite{mitroy03f,derevianko99a,derevianko01a,porsev02a,porsev03a} 
of the $C_6$, $C_8$ and $C_{10}$ dispersion coefficients suggests 
the method is able to generate coefficients that are accurate to about 
1-2$\%$.

\section{Calculation of the dispersion parameters}

All the dispersion coefficients computed in this paper were 
computed by first diagonalizing the semi-empirical Hamiltonian 
in a large mixed Laguerre type orbital or Slater type orbital 
basis set.  Next various sum rules involving oscillator 
strengths or reduced matrix elements were summed over the 
set of physical and pseudo states.

The respective formulae 
for the various coefficients are now given.  
The notation $C_n$ is used to denote the total dispersion
coefficient for a given $n$.  So for $n \ge 12$, $C_n$ is
the sum of $C^{(2)}_n$ and $C^{(4)}_n$  where $C^{(2)}_n$ 
arises from 2nd-order perturbation theory and $C^{(4)}_n$  
arises from 4th-order perturbation theory.

\subsection{The 2nd order contributions to $C_{12}$ and $C_{14}$ }

While the  higher $n$ dispersion coefficients for the H-H 
system have been determined to high precision 
\cite{thakkar88a,yan99c,mitroy05a}, 
only a few calculations of $C_{12}$ and $C_{14}$ have been
done for the alkali atoms \cite{ovsiannikov88a,patil99a}.    
The calculations of Ovsiannikov, Guilyarovski and Lopatko (OGL) 
\cite{ovsiannikov88a,mitroy05a} used an approximate expression
for the Greens function and can only be expected to be 
accurate at the 20$\%$ level.  Explicit comparisons with the 
$C_{8}$ and $C_{10}$ values of Patil and Tang (PT) 
\cite{patil99a} in references \cite{porsev03a,mitroy03f} reveal 
discrepancies of order 10$\%$ for Na$_2$, K$_2$ and 
Rb$_2$.  Recursion rules exist for estimating $C_n$ at high
$n$ from data tabulations of $C_n$ for lower values of $n$ 
\cite{thakkar88a}.
 
The polarization and dispersion parameters can be computed from 
their respective oscillator strength sum-rules which are
well known.  The 
oscillator strength, $f^{(\ell)}_{0i}$ from the ground state 
(with orbital and spin angular momentum equal zero) to the 
$i$th excited state is defined as 
\begin{equation} 
f^{(\ell)}_{0i} =  \frac {2 |\langle \psi_0 \parallel r^{\ell} 
{\bf C}^{\ell}({\bf \hat{r}} ) \parallel \psi_{i}\rangle|^2 \epsilon_{0i}}    
{(2\ell+1) }  \ .    
\label{fvaldef}
\end{equation} 
In this expression ${\bf C}^{\ell}$ is the spherical tensor of rank 
$\ell$, the Wigner-Eckart theorem is defined according to 
\cite{condon80a}, and  $\epsilon_{0i}$ is the excitation energy of the 
transition.  The sum rule for the adiabatic multipole polarizability, 
$\alpha^{(\ell)}$ is  
\begin{equation} 
\alpha^{(\ell)} = \sum_{i} \frac {f^{(\ell)}_{0i} } {\epsilon_{0i}^2} = S^{(\ell)}(-2) \ .   
\label{alphal}
\end{equation} 
The 2nd-order contributions to $C_{12}$ and $C_{14}$ are defined   
\begin{eqnarray}
C^{(2)}_{2n} &=& \sum_{\ell_i \ell_j} \frac{(2n-2)!}{2\ell_i!2\ell_j!} 
      \sum_{ij} \frac { f^{(\ell_i)}_{0i} f^{\ell_j)}_{0j} }
    {\epsilon_{0i} \epsilon_{0j}(\epsilon_{0j}+\epsilon_{0i})} \ ,      
\label{C12}
\end{eqnarray}
with $\ell_i + \ell_j + 1 = n$.  
The sum rules are a generalized sum which implicitly includes a 
sum over excitations to bound states and an integration taking 
into account excitations to continuum states.  

The sum rules involve contributions from both core and valence 
excitations.  The valence contributions were evaluated  
by diagonalizing the model Hamiltonian in a very large basis.   
Determination of the $f$-value distribution for the core   
was handled by using the properties of $f$-value sum rules 
and an approach that gives a reasonable estimate of the $f$-value 
distribution with a minimum of computation \cite{mitroy03f}.  
We use the sum-rule for the polarizability, eq.~(\ref{alphal}) 
and 
\begin{equation} 
\ell N \langle r^{2\ell-2} \rangle = \sum_{i} f^{(\ell)}_i = S^{(\ell)}(0) \ , 
\label{fsuml}
\end{equation} 
\cite{rosenthal74a,mitroy03f} to estimate an $f^{(\ell)}$-value 
distribution function of reasonable accuracy.  This expression 
reduces to the well known Thomas-Reiche-Kuhn sum rule  
for $\ell = 1$, viz 
\begin{equation} 
 N  = \sum_{i} f^{(1)}_i = S^{(1)}(0) \ .  
\label{fsum}
\end{equation} 
In these expressions $N$ is the 
total number of electrons and $\langle r^{2\ell-2} \rangle$ 
is an expectation value of the ground state wave function. 
 
The core $f^{(l)}$-value distribution was determined by assuming 
each closed sub-shell made a contribution of 
$N_i r_i^{2\ell-2}$ to the sum-rules, where $N_i$ the number 
of electrons in the sub-shell and $r_i^{2\ell-2}$ is the  
$\langle r^{2\ell-2} \rangle$ Hartree-Fock (HF) expectation value.
Next, the excitation energy for each sub-shell is set to 
the HF single particle, $\epsilon_i$ plus an 
energy shift.  The energy shift, $\Delta^{(\ell)}$ was set 
by using the core multipole polarizability, 
$\alpha^{(\ell)}_{\rm core}$ and the relation 
\begin{equation} 
\alpha^{(\ell)}_{\rm core} = \sum_{i} \frac {\ell N_i r^{2\ell-2}_i} 
{(\epsilon_{i} + \Delta^{(\ell)})^2} \ .  
\label{alphad2}
\end{equation} 
A full description of the details and core polarizabilities 
used to fix $\Delta^{(\ell)}$ for $\ell = 1, 2$ and $3$ has 
been published \cite{mitroy03f}.  
The contribution from the core was omitted 
for $\ell = 4$ and 5 since there are no reliable estimates
of these core polarizabilities and the  
$\ell r^{(2\ell-2)}_i$ weighting factors  
lead to the core becoming less important for larger $\ell$. The core
contribution to $\alpha^{(3)}$ was less than $0.2\%$ for Rb and   
most of the core contribution to $C_{2n}^{(2)}$ comes from the 
$f^{(1)}$ distribution.  

\subsection{The 3rd order and 4th order potentials}

The dispersion coefficients, $C_{11}$ and $C_{13}$, arise from
3rd-order perturbation theory 
\cite{dalgarno67a,arrighini73a,bukta74a,ovsiannikov88a,yan99c,mitroy05a}. 
The standard expressions are expressed in terms of sums
of products of reduced matrix elements.  The expressions  
derived by OGL  
\cite{ovsiannikov88a,mitroy05a} are used.  Given the complexity
of the expressions, it is not surprising there have been
relatively few calculations of these coefficients 
for any atoms.  Accurate values for $C_{11}$ through to $C_{31}$  
for the H-H system have recently been published 
\cite{yan99c,mitroy05a}. 

For the alkali atoms, the only estimates of $C_{11}$ have been obtained
from the relatively small calculations of OGL 
\cite{ovsiannikov88a} and Patil and Tang 
\cite{patil99a}.  There have been some $C_{11}$ and $C_{12}$ estimates 
for Cs 
\cite{weickenmeier85a}, but that atom is deemed too heavy to 
accurately describe with the present non-relativistic method and 
is not discussed any further here.

The dispersion coefficients, $C_{12}$ and $C_{14}$, both have
contributions that arise from 4th-order perturbation theory.
However, there have been very few calculations of the 4th
order term for any systems.  Bukta and Meath   
\cite{bukta74a} made an explicit calculation $C^{(4)}_{12}$  
for the H-H dimer.  Besides deriving a general expression   
OGL \cite{ovsiannikov88a,mitroy05a} also 
made some estimates of $C^{(4)}_{12}$ for combinations of 
hydrogen and the alkali atoms.  Finally, the present authors
made a comprehensive calculation of $C^{(4)}_{n}$ up to
$n = 32$ for the H-H interaction \cite{mitroy05a} using the 
OGL formalism. 

Although the core has not been taken explicitly into account in 
the evaluation of the 3rd and higher order dispersion parameters 
their impact is expected to be smaller than 2nd order due since 
the 3rd and 4th order sum-rules have an energy denominator 
involving the square or cube of the excitation energy.       

\begin{table}[tbh]
\caption[]{  \label{polalkali}
The higher multipole polarizabilities for the lighter alkali atoms.  
All values are in atomic units.
}
\vspace{0.1cm}
\begin{ruledtabular}
\begin{tabular}{lcccc}
Method   & \multicolumn{2}{c}{ $10^{-6}$ $\alpha^{(4)}$ } & $10^{-9}$ $\alpha^{(5)}$ & $10^{-9}$ $\alpha^{(6)}$ \\  \cline{2-3}  \cline{4-4} \cline{5-5}  
     &  present & PT \cite{patil99a} &  Present &  Present \\  \hline
Li  & 1.999 & 1.947 &    0.1551  &   17.04 \\
Na & 2.973 &  2.828   & 0.2450   &  28.53 \\
K  & 11.82  & 10.69   &  1.204    &  172.4  \\
Rb  & 16.57 & 14.49  &  1.762   & 259.6  \\
\end{tabular} 
\end{ruledtabular}
\end{table}

\section{Results of the calculation}

\subsection{Sensitivity of calculations to the ground state wave function }

One aspect of the calculations that warrants particular mention 
was the sensitivity of the higher-${\ell}$ polarizabilities 
(listed in Table \ref{polalkali}) and thus $C^{(2)}_{12}$, 
$C^{(2)}_{14}$ and $C^{(2)}_{16}$ to the representation of the ground 
state wave function.  While there are many advantages to 
representing the wave function with a linear combination of
convenient basis functions, there are some negative features.  
One of those negative features relates to the behavior at
large distances from the nucleus.  Unlike a grid based 
calculation, the correct asymptotics are not imposed and
so the large-$r$ part of the wave function, which has a 
weak influence on the total binding energy can be 
inaccurate.  

This is best illustrated by a specific calculation.  Our initial 
calculations for the $\alpha^{(4)}$ and $\alpha^{(5)}$ polarizibilities   
Na used a $3s$ wave function written as a linear combination of 
12 Slater type orbitals (STOs).  This wave function had a binding 
energy against ionization of $0.1888532$ hartree.  The 
resulting polarizabilities were  $\alpha^{(4)} = 3.46\times 10^6$ 
au and $\alpha^{(5)} = 6.42\times 10^{8}$ au.  When the STO basis 
was replaced by a large Laguerre type orbital (LTO) basis (this 
was necessary for the 
evaluation of $C_{11}$) and the energy driven to convergence 
the resulting binding energy was $0.1888549$ hartree.  However
there were dramatic changes in the polarizabilities, with the 
new values being $\alpha^{(4)} = 2.97\times 10^6$ au and 
$\alpha^{(5)}=2.45\times 10^{8}$ au (refer to Table \ref{polalkali}).  
The polarizability $\alpha^{(5)}$ decreased by a factor of about 
2.5 when the binding energy changed by $1.7\times 10^{-6}$ 
hartree!  The dispersion parameters, $C^{(2)}_{12}$  and 
$C^{(4)}_{12}$ and were also sensitive to the representation
of the ground state wave function.  The initial $C^{(2)}_{14}$ 
of $4.04\times10^{11}$ a.u. was decreased to $2.602\times10^{11}$ 
a.u. when the ground state basis was made exhaustively large.  
The impact of the basis set on the dispersion coefficients was
was not so extreme as for the polarizability but was still 
substantial. 

The $f^{(2)}$ and $f^{(3)}$ distributions
used in earlier calculations of the Na and K polarizabilities 
and dispersion coefficients \cite{mitroy03f,mitroy03g} were affected 
to a smaller extent by this problem.  The octupole polarizabilities 
were 3-4$\%$ too large for these atoms, while the effect upon 
$C_{10}$ was to make them about 1-2$\%$ too large.  The impact 
upon $\alpha^{(2)}$ and $C_8$ was an order of magnitude smaller 
but it was still discernible \cite{mitroy05a}.  Where values 
of $C_8$ and $C_{10}$ are required in the present work, the
revised values \cite{mitroy05b} are used.   

\subsection{The homo-nuclear Alkali-Alkali systems}

\begin{table*}[th]
\caption[]{  \label{Calkali}
The dispersion coefficients for homo-nuclear combinations of 
the lighter alkali atoms.  The $C^{(2)}_n$ terms have core 
contributions.   All values are in atomic units.
}
\vspace{0.1cm}
\begin{ruledtabular}
\begin{tabular}{lccccccccc}
Method   & $10^{-6}$ $C_{11}$ & $10^{-9}$ $C^{(2)}_{12}$ & $10^{-9}$ $C^{(4)}_{12}$ & 
          $10^{-9}$ $C_{13}$ & $10^{-12}$ $C^{(2)}_{14}$ & $10^{-12}$ $C^{(4)}_{14}$ &  
          $10^{-12}$ $C_{15}$ & $10^{-15}$ $C^{(2)}_{16}$ & $10^{-15}$ $C^{(4)}_{16}$ \\
\hline
\multicolumn{10}{c}{Li} \\  
Present  &  -40.44 & 0.9015  & 0.0401    & -11.05   & 0.1455 &   0.04251 & -2.873  &  0.02997 & 0.01908 \\
PT  \cite{patil99a}  & -37.36  & 0.8648    &      & -10.11   & & \\
OGL \cite{ovsiannikov88a} & -36.0   & 0.648    & 0.0433 &  -8.85   & &\\
\hline
\multicolumn{10}{c}{Na} \\  
Present   & -61.01  & 1.500  & 0.06699  & -18.90    & 0.2602 &  0.0720 & -5.368 & 0.05736  & 0.03637  \\
PT  \cite{patil99a}   & -53.32  & 1.375     &    & -16.46   & & \\
OGL \cite{ovsiannikov88a} & -39.5   &  0.917   & 0.0396 & -11.3   & & \\
\hline
\multicolumn{10}{c}{K} \\  
Present  &  -364.9  & 9.102  &  0.4640   & -147.2  & 2.000 & 0.6807 & -53.02 & 0.5544 & 0.4480 \\
PT  \cite{patil99a}   &  -312.3  & 7.749     &    & -122.8  & & \\
OGL \cite{ovsiannikov88a} & -250    & 5.430   & 0.361  &  -89.3  & & \\
\hline
\multicolumn{10}{c}{Rb} \\  
Present  & -536.2  & 14.26  &  0.6986   & -236.5  & 3.314 & 1.098 & -91.20 & 0.9656  & 0.7897  \\
PT  \cite{patil99a}   & -464.4  & 11.56    &    & -195.1  & & \\
OGL \cite{ovsiannikov88a} & -376    &  8.370   & 0.556  & -145.0  & & \\
\end{tabular} 
\end{ruledtabular}
\end{table*}

The results of the present calculations for the Li, Na, K and Rb
homo-nuclear alkali atom pairs are listed in Tables 
\ref{polalkali} and \ref{Calkali},  
and compared with the calculations of Ovsiannikov {\em et al} 
\cite{ovsiannikov88a} and Patil and Tang \cite{patil99a}.

All of the present values of $C^{(2)}_{12}$ are larger than the
PT and OGL estimates by amounts ranging from 10-50$\%$.  
This is not a concern since  
the OGL and PT estimates of $C_6$, $C_8$ and $C_{10}$ 
are also smaller than the latest data for these parameters 
\cite{porsev02a,porsev03a,mitroy03f}.  The present estimates
of $C_{11}$ and $C_{13}$ are also 10-50$\%$ larger 
than the PT and OGL estimates.     

In Table \ref{Calkali}, $C^{(2)}_{n}$ and $C^{(4)}_{n}$ are
given as separate entries.  The total dispersion parameter 
$C_{n}$ is given in Table \ref{Chet} which tabulates the
dispersion parameters for all possible atom-atom combinations. 
The 4th-order terms, $C^{(4)}_{12}$ are about 4-5$\%$ the 
size of $C^{(2)}_{12}$.  The 4th-order correction to 
$C_{12}$ is about the same size as the correction due
to the core.  The size of the 4th-order correction to
$C_{14}$ is larger, with about 25$\%$ of the 
final value of $C_{14}$ coming from $C^{(4)}_{12}$.  
For $C_{16}$, the 4th-order terms are almost as large as
the 2nd-order terms, with $C^{(4)}_{16}$ 
being 80$\%$ the size of $C^{(2)}_{16}$ 
for K$_2$ and Rb$_2$. 
  
The contribution of the core to $C^{(2)}_{12}, C^{(2)}_{14}$ 
and $C^{(2)}_{16}$ is relatively small.  The contribution is 
largest for the Rb$_2$ dimer, but even here the effect is  
4.2$\%$ for $C^{(2)}_{12}$, 3.2$\%$ for $C^{(2)}_{14}$ and 
2.5$\%$ for $C^{(2)}_{16}$.  

\subsection{Critical radius for dispersion formula}

The LeRoy radius is often used as an estimate of the
critical radius beyond which the interaction can be 
described by the use of a dispersion formula 
\cite{leroy73a,zemke94a}.  It is defined for two atoms, 
A and B 
as 
\begin{equation} 
R_{LR} = 2 \bigl( \sqrt{\langle r^2 \rangle_A} + \sqrt{\langle r^2 \rangle_B} \bigr) \ ,  
\end{equation} 
where $\langle r^2 \rangle_A$ is evaluated for the 
ground state.  The LeRoy radius for the homo-nuclear
dimers are given in Table \ref{Cradius}.      

We introduce some parameters so that the range of validity 
of the dispersion formula can be discussed in a quantitative 
manner.  First, the partial sum of the dispersion energy up to
the $n$th term is defined as 
\begin{equation}
     W_n(r) = \sum_{m=6}^{m=n} V_m(r)   \ .  
\end{equation}
One would then consider the relative size of $W_n(r)$ to 
$W_{\infty}(r)$ as a measure of the accuracy of the 
truncated dispersion potential to the exact potential.  
There are of course problems associated with the evaluation
of $W_{\infty}$, and so we identify $W_{\infty}(r)$ with 
$W_{16}(r)$.  It is natural to stop the analysis at
$n = 16$ since the contributions from 5th and 6th-order 
perturbation theory start at $n = 17$ and $n = 18$ 
respectively.  

\begin{figure}[th]
\includegraphics[width=3.5in]{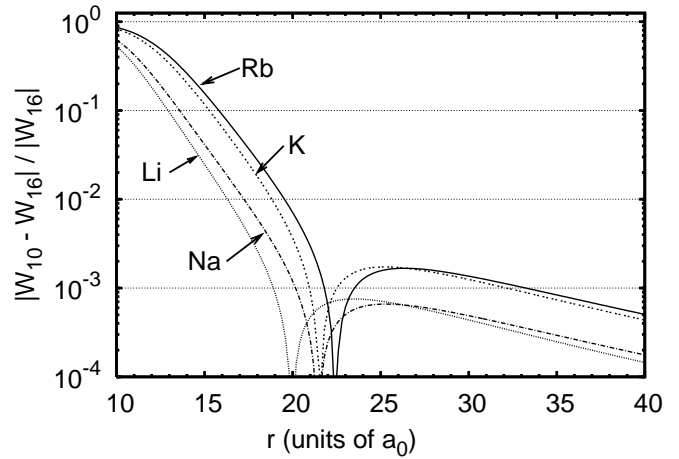}
\caption[]{ \label{W10}
The ratio of $|W_{10}(r)-W_{16}(r)|/|W_{16}(r)|$ as a function 
of $r$ (units of $a_0$) for all the possible homo-nuclear pairs. 
}
\end{figure}

Figure \ref{W10} shows the ratio 
$|W_{10}(r)-W_{16}(r)|/|W_{16}(r)|$ as a function of $r$ for 
the Li, Na, K and Rb dimers.  This ratio
is seen to be smaller than 1$\%$ for all values of $r$ 
greater than 20 a$_0$. The curves for Li$_2$ and Na$_2$ are
very similar, as are the curves for K$_2$ and Rb$_2$.  The
existence of two sets of two very similar curves is probably 
related to the fact that the lowest lying $d$-excitations for 
Li and Na involve a change in principal quantum number whereas
those for K and Rb do not.    

There is a sign change in  
$(W_{10}(r)-W_{16}(r))$ near 20 $a_0$ for all four dimers.  The 
magnitude of $W_{10}(r)$ is generally smaller than $W_{16}(r)$ 
for small separations, but for large separations $W_{10}(r)$ is 
generally larger in magnitude than $W_{16}(r)$.  This is caused 
by the repulsive $V_{11}(r)$, $V_{13}(r)$ and $V_{15}(r)$ 
interactions.

\begin{table}[th]
\caption[]{  \label{Cradius}
Various radial distances (in $a_0$) related to the accuracy 
of different order expansions of the dispersion parameters
for the homo-nuclear alkali-metal atom dimer. 
}
\vspace{0.1cm}
\begin{ruledtabular}
\begin{tabular}{lccccc}
 Atom & $\langle r^2 \rangle$ & $R_{LR}$ & $R_{6}$ & $R_{8}$ & $R_{10}$ \\ \hline 
 Li   & 17.47  & 16.72   & 77.53 &  25.92  & 16.30 \\
 Na   & 19.51  & 17.67   & 86.24  & 28.07  & 17.24 \\
 K    & 27.97  & 21.16   & 103.8  & 32.61  & 18.82 \\
 Rb   & 30.76  & 22.18   & 111.0  & 34.44  & 19.54 \\
\end{tabular} 
\end{ruledtabular}
\end{table}
 
A useful way to parameterize this information 
is to define $R_{n}$ such that it gives the smallest radius 
for which the partial dispersion energy $W_n(r)$ is 
accurate to 1$\%$.  In effect, $R_n$ is the largest $r$ solution 
of the equation  
\begin{equation}
     |W_n(r)-W_{16}(r)|  = 0.010 \times W_{16}(r)  \ \ \ n \le 16    
\end{equation}
Table \ref{Cradius} gives $R_{n}$ for the hydrogen and alkali 
homo-nuclear dimers.  The values of $R_{n}$ get larger
as the atom gets heavier.  A dispersion potential only 
involving $C_6$ would not be accurate to 1$\%$ until 
the separation distance increased to more than 100 $a_0$ 
in the case of Rb$_2$.  This distance shrinks dramatically 
with the inclusion of the $V_{8}$ and $V_{10}$ potentials.  
The size of $R_8$ indicates that a dispersion interaction 
with only $V_6$ and $V_8$ is not good enough to describe
these alkali dimers.  The $R_{10}$ parameter is smaller than 
the LeRoy radius for all systems, and so the use of $W_{10}(r)$
will be accurate to better than 1$\%$ as long as the 
internuclear separation is greater than $R_{LR}$.  

\begin{figure}[th]
\includegraphics[width=3.5in]{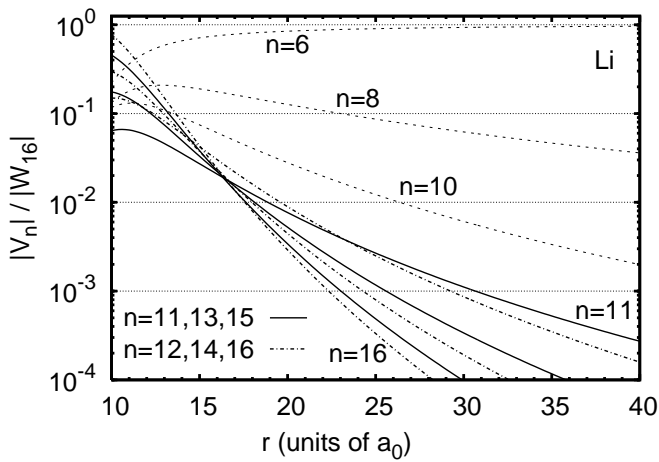}
\caption[]{ \label{LiVn}
The ratio of $|V_{n}(r)|/|W_{16}(r)|$ as a function of $r$ 
(units of $a_0$) for the lithium dimer.  The decay of the 
ratio at large $r$ is fastest for large $n$.
}
\end{figure}

\begin{figure}[th]
\includegraphics[width=3.5in]{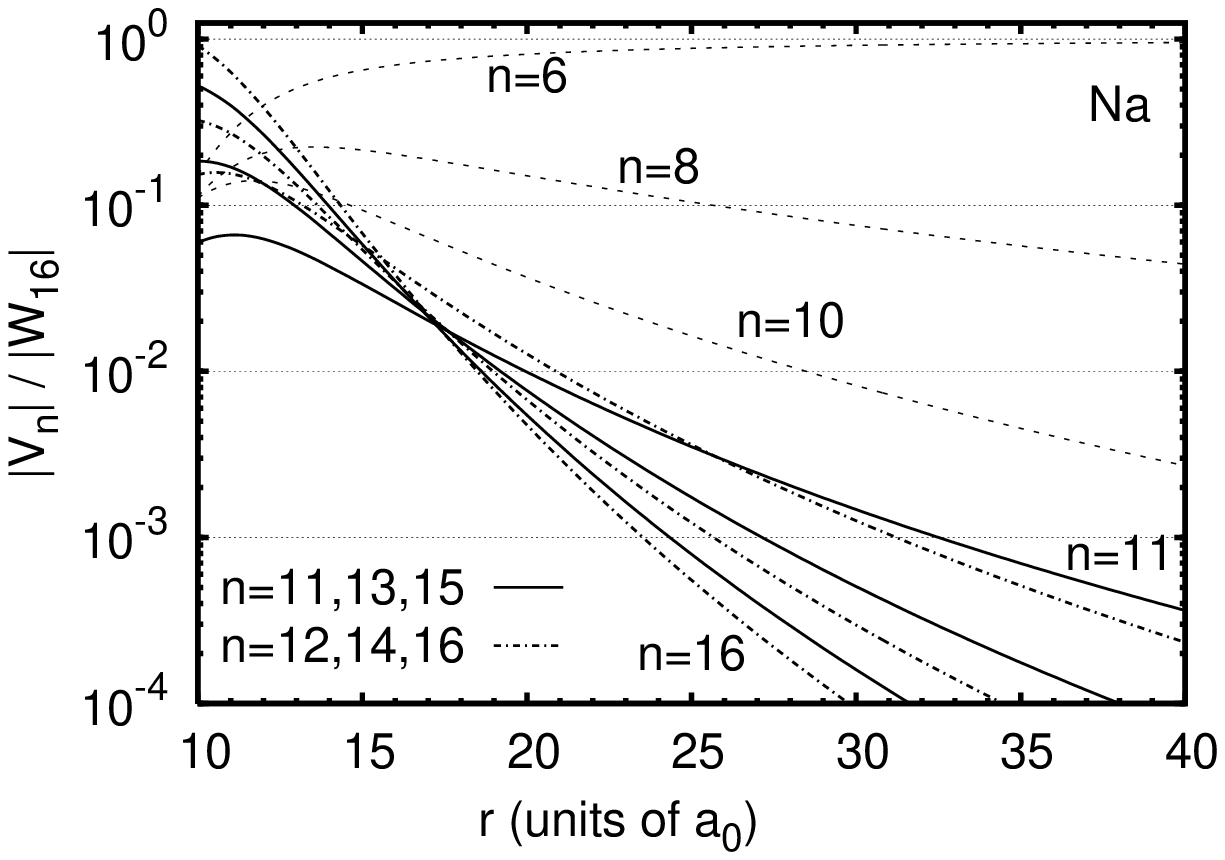}
\caption[]{ \label{NaVn}
The ratio of $|V_{n}(r)|/|W_{16}(r)|$ as a function of $r$ 
(units of $a_0$) for the sodium dimer.  The decay of the 
ratio at large $r$ is fastest for large $n$.
}
\end{figure}

\begin{figure}[th]
\includegraphics[width=3.5in]{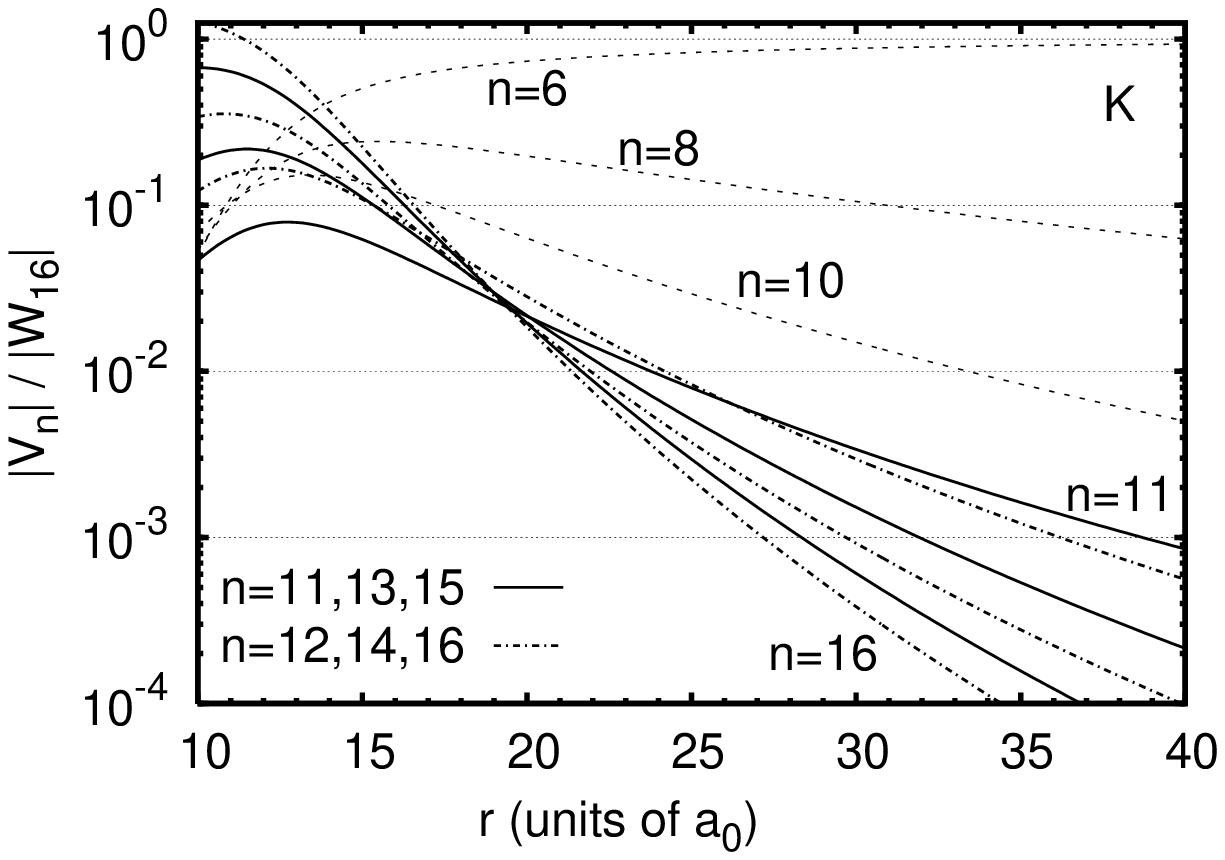}
\caption[]{ \label{KVn}
The ratio of $|V_{n}(r)|/|W_{16}(r)|$ as a function of $r$ 
(units of $a_0$) for the potassium dimer.  The decay of the 
ratio at large $r$ is fastest for large $n$.
}
\end{figure}

\begin{figure}[th]
\includegraphics[width=3.5in]{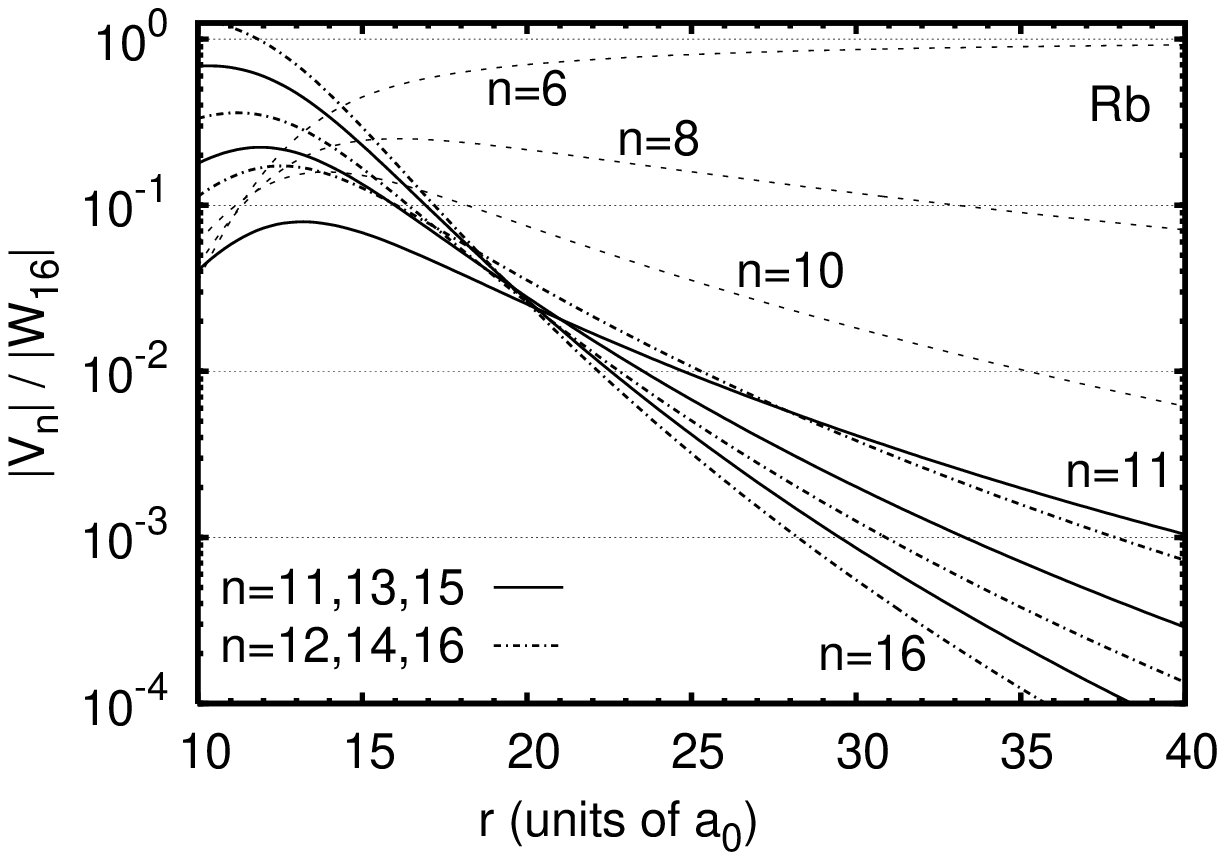}
\caption[]{ \label{RbVn}
The ratio of $|V_{n}(r)|/|W_{16}(r)|$ as a function of $r$ 
(units of $a_0$) for the rubidium dimer.  The decay of the 
ratio at large $r$ is fastest for large $n$.
}
\end{figure}

Figures \ref{LiVn}, \ref{NaVn}, \ref{KVn} and \ref{RbVn} show 
the ratio $|V_n(r)|/|W_{16}(r)|$ as a function of $r$ for all
the alkali dimers.  One of the most noticeable features
of these curves is the existence of a nexus point for 
the $V_{11}$, $V_{13}$, $V_{14}$, $V_{15}$ and $V_{16}$ 
potentials between 16-21 $a_0$ for all alkali dimers.  
This suggests that a recursion relation of the type 
$C_{n+m} = A^m C_n$ exists between the respective 
dispersion coefficients.   

The biggest correction to $W_{10}(r)$ for sufficiently 
large $r$ should be $V_{11}(r)$.   The $C_{12}$:$C_{11}$ 
ratios for Li$_2$, Na$_2$, K$_2$ and Rb$_2$ from Table 
\ref{Chet} are 23.3, 25.7, 26.2 and 27.9 respectively
and, for, $r$ less than these values it is actually 
$V_{12}(r)$ which is the largest correction to $W_{10}(r)$ 
(note; at sufficiently small $r$, $V_{13}$, $V_{14}$, 
$\ldots$ will also be larger than $V_{11}$).   

The extent to which mutual cancellations act to minimize  
the error can be gauged by adding the magnitudes of $V_n(r)$
at $r = 20$. For Na, the $\sum_{n=11}^{16} |V_n(r)|$ 
is equal to $0.047 \times W_{16}(r)$ at $r = 20 a_0$.  
But, the partial sum  $|W_{10}(r)-W_{16}(r)|$ is 
equal to $0.0012 \times W_{16}(r)$ at this radius. 
The alternating signs reduce the impact of the  
higher order terms by a factor of about 40.  
For Rb, the $\sum_{n=11}^{16} |V_n(r)|$ sum gives 
$0.168 \times W_{16}(r)$ at $r = 20$ $a_0$.  The magnitudes 
of $|V_n(r)|/|W_{10}(r)|$ are greater than 0.025 for all $n$ 
between 11 and 16.  The partial sum $|W_{10}(r)-W_{16}(r)|$ 
is equal to $0.0071 \times W_{16}(r)$ at this radius.  In this 
case, the alternating signs have reduced the impact of the 
higher order terms by a factor of more than 20.  

\begin{figure}[th]
\includegraphics[width=3.5in]{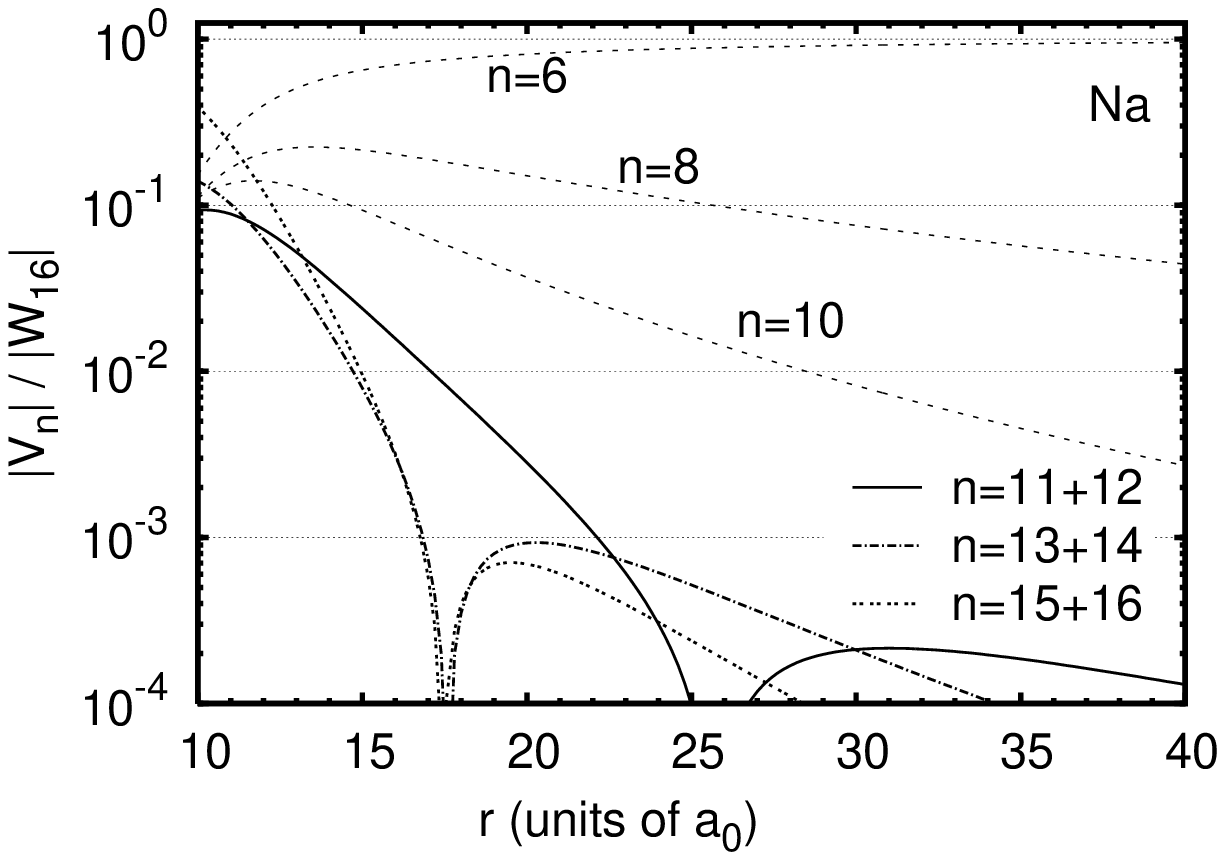}
\caption[]{ \label{Na2Vn}
The ratio of $|V_{n}(r)|/|W_{16}(r)|$ for $n =$ 6, 8 and 10 as a 
function of $r$ (units of $a_0$) for the sodium dimer.  The other 
curves show the ratio of $|V_{2n-1}(r)+V_{2n}(r)|/|W_{16}(r)|$ 
for $2n =$ 12, 14, 16. 
}
\end{figure}

\begin{figure}[th]
\includegraphics[width=3.5in]{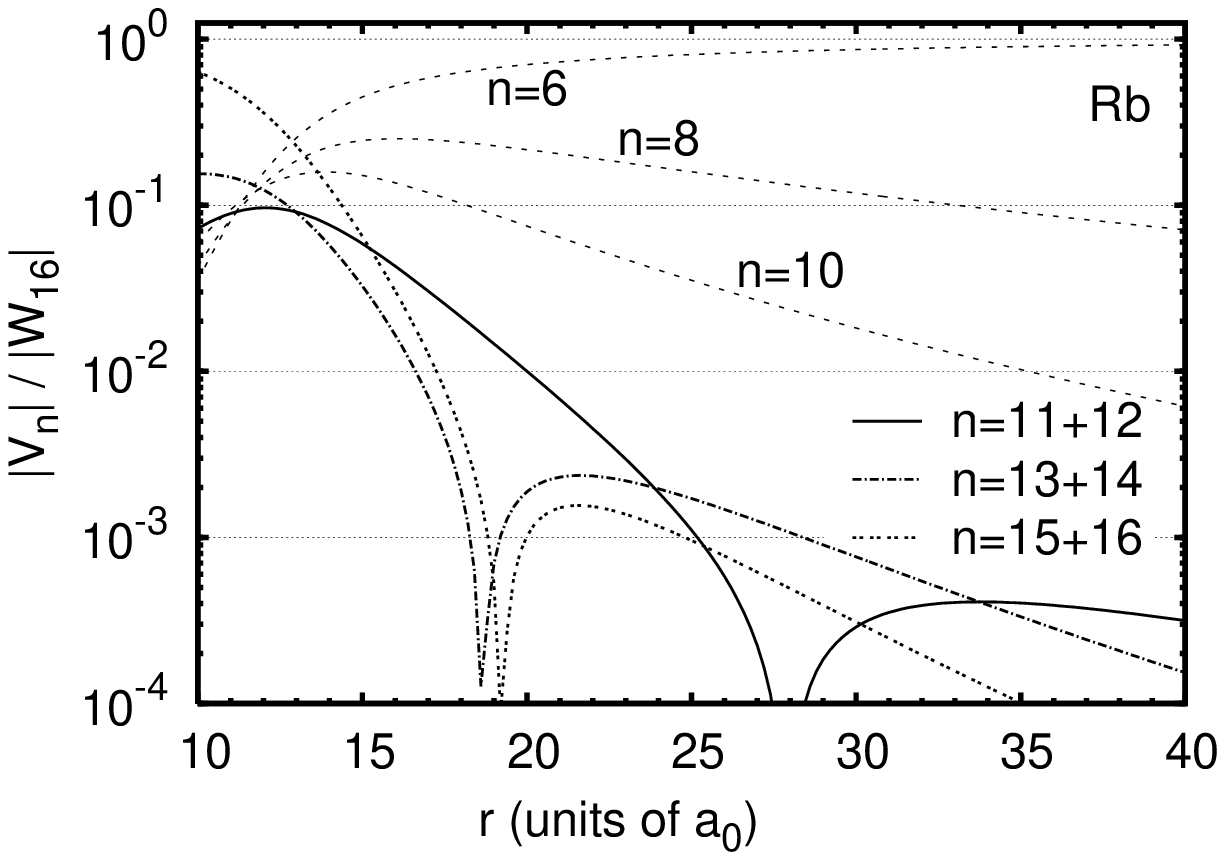}
\caption[]{ \label{Rb2Vn}
The ratio of $|V_{n}(r)|/|W_{16}(r)|$ for $n =$ 6, 8 and 10 as a 
function of $r$ (units of $a_0$) for the rubidium dimer.  The  
other curves show the ratio of $|V_{2n-1}(r)+V_{2n}(r)|/|W_{16}(r)|$ 
for $2n =$ 12, 14, 16. 
}
\end{figure}

Figures \ref{Na2Vn} and \ref{Rb2Vn} better illustrate the 
extent to which mutual cancellations result in $W_{10}$ being 
a very good approximation to the total dispersion potential 
for Na and Rb.  The combination $(V_{11}(r)+V_{12}(r))$ is
much smaller than either $V_{11}$ or $V_{12}$, and similarly 
$(V_{13}(r)+V_{14}(r))$ is smaller than $V_{13}$ or $V_{14}$ 
while $(V_{15}(r)+V_{16}(r))$ is smaller than 
$V_{15}(r)$ or $V_{16}(r)$.  The combination of
$V_{2n-1}(r) + V_{2n}(r)$ falls to 1$\%$ of $W_{16}(r)$ about
5-10 $a_0$ closer to the origin than either 
$V_{2n-1}(r)$ or $V_{2n}(r)$.

\subsubsection{The Rubidium dimer}

A recent analysis of three complementary experiments
by van Kempen {\em et al} \cite{vankempen02a} 
for the Rb$_2$ dimer resulted in experimental estimates  
of the dispersion parameters $C_{6}$, $C_{8}$, $C_{10}$
and $C_{11}$ and gave estimates of the scattering length 
to a precision of about $1\%$ for Rb$^{87}$.  

The present work suggests that their estimates of  
$C_{6}$, $C_{8}$ and $C_{10}$ may be well founded and 
our previous calculations agreed with the experimental
values to an accuracy of 2$\%$ \cite{mitroy03f}.  
However, it is probable that their attempt to determine 
$C_{11}$ was overly ambitious.  The present calculation 
gave  $C_{11} = -5.362 \times 10^{9}$ au which lies outside 
the van Kempen estimate of 
$C_{11} = (-8.6 \pm 0.17) \times 10^{9}$ au.  We believe 
this difference is a consequence of assumptions made by 
van Kempen {\em et al} when they performed the fit 
to extract the dispersion coefficient.

First, their least squares fit relied on a value of $C_{12}$ 
of $1.19\times 10^{10}$ au computed by Patil and Tang 
\cite{patil97a}.  This estimate of  $C_{12}$ is about $25\%$ 
smaller than the present 2nd plus 4th order $C_{12}$ 
of $1.496 \times 10^{10}$ au.  Second, their estimate is 
derived from the energies of bound states that have an 
outer turning radius of about 20 $a_0$.  At this radius, 
$V_{12}$ is actually larger than $V_{11}$.  Furthermore, 
the higher order terms, $V_{13}$, $V_{14}$, $V_{15}$ and 
$V_{16}$ were all slightly larger than $V_{11}$ at a radius 
of 20 $a_0$!  However, as has been noted, the
alternating signs of the successive terms results in 
a considerable degree of cancellation.  Nevertheless, 
one can conclude that the van Kempen estimate of 
$C_{11}$ is sensitive to the accuracy and presence of the 
high $n$ terms in the dispersion interaction. 

\subsection{The hetero-nuclear systems including H}

The $n>10$ dispersion parameters for all possible combinations 
of H, Li, Na, K and Rb are given in Table \ref{Chet}.  The
radial matrix elements for hydrogen were those of the $N=15$ 
calculation used in an earlier calculation of the H-H dispersion 
parameters \cite{mitroy05a}.   The $C_{11}$ and $C_{12}$ parameters 
reported by OGL \cite{ovsiannikov88a} and PT \cite{patil99a} are 
not listed since the present calculations are more sophisticated.

\begin{table*}[th]
\caption[]{  \label{Chet}
The dispersion coefficients, $C_{11} \rightarrow C_{16}$ for all 
the possible interacting pairs formed by hydrogen and the alkali 
atoms.  The $C_{12}$, $C_{14}$ and $C_{16}$ coefficients have  
contributions from both
2nd- and 4th-order perturbation theory. All values are in atomic units.
}
\vspace{0.1cm}
\begin{ruledtabular}
\begin{tabular}{lcccccc}
System   & $C_{11}$ & $C_{12}$ & $C_{13}$  & $C_{14}$ & $C_{15}$ & $C_{16}$  \\ \hline
H-H \cite{mitroy05a}  & -3474.9 & $1.2273\times10^5$  & -$3.2699 \times 10^5$  &   $6.3617\times 10^6$  &  
       -$2.8396\times10^{7}$ &  $4.4121\times10^{8}$  \\ 
\\  
        & $10^{-6}$ $C_{11}$ & $10^{-6}$ $C_{12}$ & $10^{-9}$ $C_{13}$ & $10^{-9}$ $C_{14}$ & 
       $10^{-9}$ $C_{15}$  &  $10^{-12}$ $C_{16}$  \\ \hline
H-Li   &  -0.2251  & 21.17   & -0.04752  & 2.735  &  -9.438 &  0.4593  \\
H-Na   &  -0.2760  & 29.43   & -0.06269  & 4.030  & -13.25  &  0.7160   \\
H-K    &  -0.5636  & 92.69   & -0.1635   & 15.68  & -43.10  &  3.409  \\
H-Rb   &  -0.6619  & 12.34   & -0.2064   & 21.86  & -57.57 &  4.936  \\     
\\ 
        & $10^{-6}$ $C_{11}$ & $10^{-9}$ $C_{12}$ & $10^{-9}$ $C_{13}$ & $10^{-12}$ $C_{14}$ & 
       $10^{-12}$ $C_{15}$  &  $10^{-13}$ $C_{16}$  \\ \hline
Li-Li  &  -40.44  & 0.9417    & -11.05   &  0.1880  &  -2.873  &  4.906  \\
Li-Na  &  -49.63  & 1.221     & -14.48   &  0.2512  &  -3.937  &  6.819  \\
Li-K   &  -119.4  & 3.216     & -40.67   &  0.7685  &  -12.68  &  24.08  \\
Li-Rb  &  -143.9  & 4.134     & -51.84   &  1.018   &  -16.84  &  32.90  \\    
Na-Na  &  -61.01  & 1.567     & -18.90   &  0.3323  &  -5.368  &  9.373  \\
Na-K   &  -146.5  & 4.015     & -52.60   &  0.9834  &  -17.04  &  31.97  \\
Na-Rb  &  -176.5  & 5.125     & -66.85   &  1.292   &  -22.54  &  43.32  \\    
K-K    & -364.9   & 9.567     & -147.2   &  2.681  &  -53.02  &  100.24  \\
K-Rb   & -442.1   &  11.99    & -186.7   &  3.449  &  -69.63  & 133.0   \\    
Rb-Rb  & -536.2   &  14.96    & -236.5   &  4.412  &  -91.20  & 175.5   \\
\end{tabular} 
\end{ruledtabular}
\end{table*}

\section{Conclusions}

The complete set of dispersion parameters up to $n = 16$ has 
been computed for all combinations of hydrogen and the alkali 
atoms up to rubidium.  The relative importance of the dispersion
potentials $V_{n}(r)$ increases as the atoms get heavier.  
It was found that the dispersion energy given by the 
first 3 terms of eq.~(\ref{vdw1}) is accurate to 1$\%$ whenever 
$R > 20$ $a_0$.  This degree of accuracy at relatively small 
internuclear separations comes from a fortuitous cancellation 
between the terms with $n > 10$ in the dispersion energy.  The 
3rd-order potentials, $V_{11}$, $V_{13}$ and $V_{15}$ are repulsive, 
while the even terms, $V_{12}$, $V_{14}$ and $V_{16}$ are 
attractive.  The 4th-order contribution to $C_{16}$ was almost
as large as the 2nd-order contribution to $C_{16}$.

Whether terms in the dispersion interaction with $n > 10$ 
are important in the description of alkali dimers is 
essentially a question about whether the dispersion
interaction has to be known to a precision of 1.0$\%$ or 
0.1$\%$.  It is clear that additional terms going beyond 
$C_{10}$ should be introduced in pairs.  There is 
no point in including the $V_{11}$ potential without also 
including the $V_{12}$ potential.  Indeed, inclusion of 
just one member of the ($V_{11}$,$V_{12}$) would most
likely degrade rather than improve the accuracy of the dispersion
potential.   Given that the $C_6$ parameter has been calculated 
to a precision of better than 1$\%$ for most alkali systems 
\cite{mitroy03f,porsev02a,porsev03a}, usage of a dispersion
interaction involving the ($V_{11},V_{12})$ potentials 
may be warranted.  

There have been a couple of experimental investigations of 
alkali-dimer potentials that have included dispersion forces
with $n > 10$.  The value of $C_{11}$ for the Rb dimer has 
been given by van Kempen {\em et al} \cite{vankempen02a}. 
However, this value of $C_{11}$ is most likely model 
dependent for reasons discussed earlier.  Consideration of 
the $V_{12}(r)$ potential has also occurred in analyses of 
the spectrum of the Cs dimer \cite{amiot02b,vanhaecke04a}.
In this case, the $V_{11}(r)$ potential was omitted so 
it is doubtful whether the inclusion of $V_{12}(r)$ was 
justified in this case. 

\section{Acknowledgments}

The authors would like to thank Mr J C Nou and Mr C Hoffman  
of CDU for workstation support.  


\end{document}